\documentclass[review,sort&compress]{elsarticle}

\usepackage{lineno,hyperref}
	\hypersetup{colorlinks=true,pdfborder=0 0 0}
\usepackage{subcaption}
\usepackage{amsmath}
\usepackage{amssymb}
\usepackage{gensymb}
\usepackage{cleveref}
\usepackage{siunitx}
\usepackage[dvipsnames]{xcolor}
\modulolinenumbers[5]

\RequirePackage[normalem]{ulem} 
\RequirePackage{color}\definecolor{RED}{rgb}{1,0,0}\definecolor{BLUE}{rgb}{0,0,1} 

\newcommand{\norm}[1]{\left\lVert#1\right\rVert}

\journal{Comp.\ Meth.\ Appl.\ Mech.\ Eng.}

\begin{document}  

\begin{frontmatter}

\title{Materials knowledge system for nonlinear composites}

\author[ucsb-me,ucsb-mat]{Marat I. Latypov}
\ead{latmarat@ucsb.edu}
\author[labex]{Laszlo S. Toth}
\ead{laszlo.toth@univ-lorraine.fr}
\author[gatech-me,gatech-cs]{Surya R. Kalidindi \corref{cor1}}
\ead{surya.kalidindi@me.gatech.edu}

\cortext[cor1]{Corresponding author}

\address[ucsb-me]{Mechanical Engineering Department, University of
California, Santa Barbara, USA}
\address[ucsb-mat]{Materials Department, University of
California, Santa Barbara, USA}
\address[labex]{Laboratory of Excellence on Design of Alloy Metals for low mAss Structures and Laboratoire d'\'{E}tude des Microstructures et de M\'{e}canique des Mat\'{e}riaux (LEM3), Universit\'{e} de Lorraine, CNRS, Arts et M\'{e}tiers ParisTech, LEM3, 57000 Metz, France}
\address[gatech-me]{Woodruff School of Mechanical Engineering, Georgia Institute of
Technology, Atlanta, USA}
\address[gatech-cs]{School of Computational Science and Engineering, Georgia Institute of
Technology, Atlanta, USA}

\begin{abstract}
In this contribution, we present a new Materials Knowledge System framework for microstructure-sensitive predictions of effective stress--strain responses in composite materials. The model is developed for composites with a wide range of combinations of strain hardening laws and topologies of the constituents. The theoretical foundation of the model is inspired by statistical continuum theories, leveraged by mean-field approximation of self-consistent models, and calibrated to data obtained from micromechanical finite element simulations. The model also relies on newly formulated data-driven linkages between micromechanical responses (phase-average strain rates and effective strength) and microstructure as well as strength contrast of the constituents.  The paper describes in detail the theoretical development of the model, its implementation into an efficient computational plasticity framework, calibration of the linkages, and demonstration of the model predictions on two-phase composites with isotropic constituents exhibiting linear and power-law strain hardening laws. It is shown that the model reproduces finite element results reasonably well with significant savings of the computational cost. 
\end{abstract}


\end{frontmatter}

\section{Introduction}\label{introduction}
\label{sec:intro}

Physics-based modeling of the material constitutive response is a key tool in the efforts aimed at optimizing the performance characteristics of engineered components designed to meet the stringent demands of advanced technologies. The main challenge comes from the need to incorporate the rich details of the material internal structure over a hierarchy of length scales spanning from the atomistic to the sample scales. Significant advances have been made in understanding the governing physics at selected length scales in many material classes (e.g., \cite{Roters2010,Rickman1997,Xu2018}). However, to benefit from these advances in integrated material--product optimization \cite{McDowell2010}, one needs a computationally efficient scale-bridging framework that accounts faithfully for the myriad details of the material microstructure hierarchy in modeling the overall material constitutive response.

Composite theories provide the theoretical foundations for formulating the overall response of the material based on the details of its internal structure. These theories address both localization (passing information from higher to lower length scales) and homogenization (passing information from lower to higher length scales) problems \cite{Nemat2013,Suquet1987,Matous2017} central to scale-bridging in multiscale modeling. Most of these theories were initially developed for linear elasticity with some of them extended to more challenging nonlinear phenomena such as viscoplasticity \cite{Castaneda1991,Garmestani1998}.

\emph{Mean-field theories} are often employed as the method of choice in multiscale modeling owing to their simplicity and modest computational cost. The central assumption of the original mean-field theories is that strain (rate) and stress fields within the constituents are uniform \cite{Eshelby1957,Hutchinson1976,Stringfellow1991}. In the simplest mean-field models, selected microscale fields are further assumed to be uniform throughout the composite, i.e., equal in each of the constituents in the composite \cite{Voigt1928,Reuss1929}. In the self-consistent models (e.g., \cite{Kroner1961,Hill1965,Budiansky1965,Hutchinson1976,Stringfellow1991,Molinari1994}), the mean-field response in each constituent is approximated using the Eshelby solution \cite{Eshelby1957} to an elliptical inclusion in an infinite medium. Recent progress in self-consistent theories allows for resolving stress and strain heterogeneities within the constituents \cite{Castaneda1991,Idiart2006,Castelnau2006,Lebensohn2007}. While the mean-field theoretical framework is attractive due to the modest computational requirements, the material microstructure is often captured using relatively simple descriptors. For example, in Reuss and Voigt models, the microstructure of a composite material is described only by the volume fractions of the constituents \cite{Kroner1978}. Such models are limited to providing bounds on the effective response of composites with complex heterogeneous microstructures \cite{Adams2012,Adams1998,Adams2005,Torquato1991}. These bounds are often widely separated which gives limited guidance in addressing the inverse problems encountered in microstructure optimization and design \cite{Fullwood2010}. Self-consistent models allow for incorporation of richer details of the microstructure such as the aspect ratios of ellipsoidal constituents (particles or grains in polycrystal) \cite{Lebensohn1993,Tome1999,Nebozhyn2001} as well as their morphological orientations \cite{Lebensohn1993}, which improves the estimates of the effective properties. Most of these models still do not account for the highly complex details of the 3-D spatial arrangement of the constituents measured in actual microstructures.

\emph{Statistical continuum theories} aim to address the deficiencies of the mean-field theories described above. These theories were originally devised and introduced by Brown \cite{Brown1955} with the goal of incorporating all the details of the material microstructure. The statistical continuum theories have been advanced through the works of Kr{\"o}ner \cite{Kroner1977,Kroner1972}, Torquato \cite{Torquato1997,Torquato1991,Torquato2013}, and Adams et al. \cite{Adams2012,Adams1998,Adams2005,Adams1989,Garmestani1998}. In this class of composite theories, the microstructure information is accounted through the formalism of $n$-point spatial correlations, combined with the use of the Green's functions. Although the mathematical foundations of these theories are rigorous, their practical adoption has encountered significant challenges: (i) the mathematical descriptions of the $n$-point spatial correlations recovered from real material samples are unwieldy \cite{Adams1989,Adams1998} and do not fit any simple mathematical functions, (ii) the relevant Green's functions capturing the controlling physics in a wide variety of materials phenomena are not easily established, and (iii) the series solutions obtained for the effective properties in this approach have not been found to be unconditionally convergent \cite{Binci2008}. As a result, the full potential of the statistical continuum theories has not yet been realized. 

\emph{Computational homogenization} approaches have become a viable alternative to the classical composite theories with the major advances made in recent years in the field of computational mechanics \cite{Segurado2018}. Computational homogenization includes numerical techniques such as the micromechanical finite element models (e.g., \cite{Gilormini1987,Ghosh1995,Segurado2002,Geers2010,Bosco2014,Brands2015,Latypov2016}) and the FFT-based methods \cite{Moulinec1998,Michel1999,Lebensohn2001,Eisenlohr2013}. More recent developments have included the variational asymptotic method for unit cell homogenization \cite{Yu2009}. The main advantage of the numerical methods is that many details of the actual microstructure can be accounted explicitly in the simulation. In these approaches, one usually incorporates the microstructure information by meshing or voxelizing the representative volume, and prescribes appropriate laws to capture the constitutive behavior of the individual microscale constituents present in the composite. Computational approaches therefore allow one to obtain the micromechanical fields of interest (and hence effective responses) for a variety of microstructures and constitutive laws. These numerical methods are typically quite demanding in terms of the required computational resources, especially when compared to the analytical models discussed earlier. The high computational cost of the numerical methods significantly limits their broader adoption and utilization in multiscale modeling and design \cite{Kalidindi2015,Panchal2013}. 

Multiscale modeling and design efforts could clearly benefit from a new approach that can potentially combine the formulation rigor of the statistical continuum theories, computational efficiency of the simplest mean-field models, and the versatility of the numerical methods, thereby overcoming the existing trade-offs made between the accuracy and the computational cost. An example of prior work in this direction can be seen in self-consistent clustering analysis for inelastic heterogeneous materials proposed by Liu et al.\ \cite{Liu2016}. Their two-stage analysis consists of (i) offline numerical simulation of elastic deformation followed by clustering of material points according to their elastic strain localization and (ii) self-consistent homogenization of the clusters during inelastic deformation. Another new framework based on emerging data science strategies -- Materials Knowledge Systems (MKS) \cite{Kalidindi2015,Kalidindi2015r} -- has also shown potential for addressing the trade-offs described earlier. Specifically, MKS demonstrated a combination of high accuracy and modest computation cost in addressing both localization and homogenization problems \cite{Yabansu2014,Gupta2015,Latypov2017,Paulson2017}. The central idea underlying the MKS approach is the calibration of the Green's function-based convolution kernels in the series expansions employed in the statistical continuum theories to the numerical datasets produced by micromechanical finite element simulations on ensembles of digitally created microstructure exemplars \cite{Gupta2015,Latypov2017,Paulson2017}.

In prior work, the MKS homogenization approach has been demonstrated for predictions of the effective elastic stiffness and initial yield strength of composites \cite{Gupta2015,Latypov2017,Paulson2017}. However, many applications in integrated material--product design demand computationally efficient predictions of the stress--strain response (e.g., tensile stress--strain curve) beyond the elastic stiffness and the yield strength. Such predictions are especially critical when the manufacture of the component involves deformation processing operations. The major challenge in modeling the stress--strain response of the composite comes from the need to allow for arbitrary (typically nonlinear) strain hardening laws of the constituents. In the current MKS protocols, changing hardening laws would require recalibration of the models for each new set of hardening parameters of the constituents. Consequently, the prediction of a single stress--strain curve of a composite with constituents exhibiting different strain hardening behaviors would demand large sets of calibration data requiring finite element simulations covering the range of strength contrast ratios that can potentially be observed between the phases during deformation of the composite. 

Our goal in this work is to develop and demonstrate a new MKS approach for microstructure-sensitive predictions of nonlinear stress--strain responses in composite materials made from viscoplastic constituents exhibiting arbitrary strain hardening laws. Towards this end, we first formulate a clever strategy for predicting microstructure-sensitive stress--strain responses that do not require recalibration for every new set of strain hardening laws employed for the constituents. We then integrate these MKS linkages with the classical mean-field approximation in a computationally efficient manner for predicting the stress--strain responses of a broad range of potential microstructures and choices of microscale constituents. To our knowledge, such a versatile microstructure-sensitive framework has not been reported in prior literature on composite materials. 

\section{Theoretical Development}
\label{sec:mdl-dev}

\subsection{Problem statement}
\label{sec:problem}

In this study, we specifically focus on the effective mechanical behavior of composite (or multiphase) materials made of two isotropic rigid-viscoplastic constituents (or phases), and their equivalent stress--equivalent plastic strain relationships. This is seen as the first step in the development of the new reduced-order modeling framework envisioned in this work, and will serve as the foundation for future enhancements to include more constituent phases and more complex constitutive laws.

The spatial distribution of the constituent phases in the composite that decides the effective response of interest is assumed to be adequately captured in a three-dimensional representative volume element (RVE) occupying volume $V$. The microstructure of such RVEs can be uniquely described by the microstructure function, $m(\beta,\mathbf{x})$ \cite{Adams2005,Niezgoda2011}, where $\beta$ denotes the local state, which, for composites, simply refers to the phase label. As an example, for a two-phase composite, $\beta \in \{1,2\}$. The microstructure function can be interpreted as the spatially resolved volume fraction of the constituents at every material point, $\mathbf{x} \in V$ \cite{Adams2005,Niezgoda2011}. The two constituents, thus labeled as \texttt{phase 1} (e.g., matrix) and \texttt{phase 2} (e.g., reinforcement), occupy subvolumes $V_1$ and $V_2$ of the RVE.

In this work, we restrict our attention to composites with isotropic phases, whose plasticity can be described by $\text{J}_2$-based associated flow rule. Within the $\text{J}_2$ plasticity theory, the internal state at each material point, $\mathbf{x}$, can be captured by a single state variable, $g$, that describes the local resistance to plastic deformation (henceforth referred as strength) \cite{Weber1990}. Strength, as the internal state variable, is expected to evolve during deformation, which reflects the evolution of the underlying substructure (e.g., dislocation density and configuration) and produces the strain hardening phenomenon. The strain hardening can be expressed for isotropic materials in a general form as \cite{Weber1990}

\begin{align} \label{eqn:hard_gen}
\dot{g} = h(g)\dot{\epsilon},
\end{align}

\noindent where $\dot{g}$ is time rate of change in strength, $h(g)$ is a material-specific hardening law, typically a nonlinear function of strength $g$, $\dot{\epsilon}$ is the equivalent plastic strain rate defined as $\dot{\epsilon} = \sqrt{\frac{2}{3} \norm{\mathbf{D}}}$ with $\mathbf{D}$ denoting the symmetric second-rank strain rate tensor and $\norm{\cdot}$ denoting the norm of a tensor. 

In most engineered composites, the constituents are selected with different combinations of mechanical properties so that $g_1 \ne g_2$ and $h_1 \ne h_2$. Consequently, even if the initial spatial distribution of strength, $g(\mathbf{x})$, is uniform within the phases, i.e.,  

\begin{equation} 
  g(\mathbf{x}) = 
\begin{cases}
g_1, \text{if } \mathbf{x} \in V_1, \\
g_2, \text{if } \mathbf{x} \in V_2, 
\end{cases}
  \label{eqn:s_dist}
\end{equation} 

\noindent the strain rate field in the RVE, $\dot{\epsilon}(\mathbf{x})$, will become heterogeneous upon a macroscopically imposed deformation, $\dot{\bar{\epsilon}}$. Consequently, each material point will undergo its own local deformation, which in turn will lead to non-uniform spatial distribution of strength, $g(\mathbf{x})$, according to \Cref{eqn:hard_gen}, at any time during any imposed loading history. 

Prediction of the stress--strain response of the composite therefore generally requires solution of the localization problem, i.e., obtaining $\dot{\epsilon}(\mathbf{x})$ for the given macroscopic deformation, $\dot{\bar{\epsilon}}$. Once the strain rate field is known, the strength can be updated locally using \Cref{eqn:hard_gen} at each material point. Having the spatial distribution of strength, one needs to address the homogenization problem to obtain the effective or overall strength of the composite, $\bar{g}$, which in turn provides for effective stress response of the composite, $\bar{\sigma}$, e.g., $\bar{\sigma}=\bar{g}$ for rate-independent materials. Our problem statement therefore requires us to address (i) a microstructure-sensitive solution to the localization problem, (ii) local update to account for the strain hardening, and (iii) a microstructure-sensitive solution to the homogenization problem. 

\subsection{Mean-field approximation of localization and strain hardening}
\label{sec:mean-field}

Complete solution to the localization problem, i.e., prediction of the local strain rate field for a given macroscopically imposed strain rate generally requires employing either the statistical continuum theories (still under development) or the computationally demanding numerical methods. Prior work employing self-consistent models (e.g., \cite{Stringfellow1991,Stringfellow1992}) has demonstrated the remarkable utility of the simplified mean-field approaches to micromechanics by tracking only phase-averages, i.e., values obtained by averaging micromechanical fields over the phase volumes. In this work, we are specifically interested the phase-average equivalent strain rates, $\dot{\epsilon}_\beta$, which we define as:

\begin{align} \label{eqn:eps_ph}
\dot{\epsilon}_\beta = \int_{V_\beta} \dot{\epsilon} dV,
\end{align}

The mean-field approximation results in significant simplification of the localization problem by requiring solutions for only phase-averages, in this case, for equivalent strain rates, $\dot{\epsilon}_\beta$, or equivalently, strain-rate partitioning ratios \cite{Stringfellow1991}: 

\begin{align} \label{eqn:chi}
\chi_\beta = \dot{\epsilon}_\beta/\dot{\bar{\epsilon}}.
\end{align}

\noindent where $\dot{\bar{\epsilon}}$ is the macroscopically imposed equivalent strain rate. Furthermore, consistent with this approach, the problem of capturing strain hardening (\Cref{eqn:hard_gen}) can also be reduced to its mean-field approximation, which will require updating only the phase-average strength, $g_\beta$:

\begin{align} \label{eqn:s_ph}
g_\beta  =\int_{V_\beta} g dV.
\end{align}

\noindent Consideration of the phase-average strain rates, $\dot{\epsilon}_\beta$, and strengths, $g_\beta$ allows us to write the following mean-field approximation of \Cref{eqn:hard_gen}: 

\begin{align} \label{eqn:hard_ph}
\dot{g}_\beta = h_\beta(g_\beta)\dot{\epsilon}_\beta,
\end{align}

\noindent Accordingly, the problem of obtaining effective strength of the composite, $\bar{g}$, reduces from homogenizing the highly complex strength field $g(\mathbf{x})$ to homogenizing a two-phase structure with average strengths of $g_1$ and $g_2$.

The reader should note that our definition of phase-average equivalent strain rate written in \Cref{eqn:eps_ph} departs from the usual convention in micromechanics. Specifically, we integrate von Mises equivalent strain rates over the phase volume, whereas the usual approach is to volume-average the strain rate tensor components and then find the von Mises equivalent of the volume-averaged strain rate tensor. The phase-average equivalent strain rate was adopted because our studies have shown that volume-averaging of equivalent strain rates captures more accurately the strain hardening of the isotropic phases when strain rates and strengths are assumed to be uniform in the phase volumes. Adopting this definition (as opposed to equivalent of the averaged strain rate tensor) mitigates the mismatch in the full-field and mean-field strain hardening responses, which originates from intraphase heterogeneities of strain and local state variables \cite{Castelnau2006}. With these definitions, generally $\sum_\beta f_\beta\chi_\beta \ne 1$ and  $\dot{\bar{\epsilon}} \ne \sum_\beta f_\beta\dot{\epsilon}_\beta$ with $f_\beta$ denoting the volume fraction of the $\beta^\text{th}$ phase. We also note that this deviation is applied only to the strain rate tensor and not the stress tensor. For the stress, the macroscopic stress tensor is computed by volume-averaging, and then the Mises definition is applied to obtain the macroscopic equivalent stress.

While the classical models based on the mean-field approximation (\Cref{eqn:hard_ph}) proved to be useful for certain cases \cite{Stringfellow1991,Stringfellow1992}, their main deficiency is that the predictions of the strain-rate partitioning ratios ($\chi_\beta$, and hence, phase-average strain rates, $\dot{\epsilon}_\beta$) as well as the effective strength, $\bar{g}$, are based on limited microstructure descriptors. In this work, we will adopt the mean-field approximation of localization and strain hardening, but address strain-rate partitioning and homogenization problems using the MKS approach to account for the higher-order details of the microstructure.

\subsection{Current MKS approach}

As introduced in \Cref{sec:intro}, the MKS approach provides a data-science framework for practical implementation of statistical continuum theories, i.e.,  microstructure-sensitive solutions for micromechanics. The key components of the MKS approach to homogenization problems include (i) discretization of the microstructure and spatial statistics, (ii) low-dimensional representation of spatial correlations, and (iii) their incorporation into a reduced-order microstructure--property linkage. Specifically, the property dependence on the microstructure is expressed as a series expansions analogous to the statistical continuum theories, but in terms of the low-dimensional representations of the spatial statistics. Such a microstructure--property linkage for a general effective property, $\bar{p}$, can be expressed as a multivariate polynomial function \cite{Gupta2015,Paulson2017,Latypov2017}:

\begin{align} \label{eqn:poly1}
    \bar{p} = \sum_q A_q \alpha^q,
\end{align}

\noindent where $\alpha^q = \alpha_1^{q_1}\alpha_2^{q_2} \ldots \alpha_R^{q_R}$ are power products (monomials) of the scores (i.e., microstructure features identified in principal component analyses (PCA) of the spatial statistics) $\alpha = \{\alpha_1, \alpha_2,\ldots,\alpha_R \}$, and $A = \{A_0,A_1,\ldots, A_R\}$ are influence coefficients capturing the microstructure--property relationship. Multi-index $q$ is an array of exponents, $q = \{q_1,q_2,\ldots,q_R\}$, each element of which is a non-negative integer allowed to vary from 0 to the selected maximum degree, $Q$ (i.e., $\ q \in [0,Q]$). The polynomial function in \Cref{eqn:poly1} also contains a free constant, $A_0$, which corresponds to the array of zero exponents $\{q_1,q_2,\ldots,q_R\}=\{0,0,\ldots,0\}$. For example, a polynomial function corresponding to sets $\alpha = \{\alpha_1,\alpha_2\}$ and $q = \{1,2\}$ would be written as $\bar{p} = A_0 + A_1\alpha_2 + A_2\alpha_2^2 + A_3\alpha_1 + A_4\alpha_1\alpha_2 + A_5\alpha_1\alpha_2^2$.

In the formulation presented above, the low-dimensional representation of the microstructure was obtained using principal component analysis (PCA) \cite{Niezgoda2013,Bro2014,Latypov2018} so that the terms $\alpha$ represent the PC scores. As a consequence, the vector array of spatial statistics, $\mathbf{f}^{(k)}$, for any $k^\text{th}$ microstructure in an ensemble can be represented by

\begin{align} \label{eqn:pca}
    \mathbf{f}^{(k)} \approx \sum_{j=1}^{R} \alpha_j^{(k)} \boldsymbol{\phi}_{j} + \langle \mathbf{f} \rangle, 
\end{align}

\noindent where $\langle \mathbf{f} \rangle$ is the average of the spatial statistics for all microstructures in the ensemble studied, $\boldsymbol{\phi}_{j}$ is the corresponding set of vector bases of the PC space, and $R$ is the truncation level selected to satisfy \Cref{eqn:pca} to a good approximation. Dimensionality reduction is attained when $R$ is significantly smaller than the number of spatial statistics (i.e., the length of $\mathbf{f}^{(k)}$), which is usually the case for the class of applications considered in this work \cite{Niezgoda2008}. While the formulation in \Cref{eqn:poly1,eqn:pca} permits incorporation of a set of $n$-point correlation functions of infinitely high order, prior work on MKS model development showed that consideration of 2-point correlations functions (also simply called 2-point statistics) suffice for capturing microstructure effects on a micromechanical response of composites with low to moderate contrasts \cite{Fast2011,Latypov2017}. An expression for 2-point statistics for a discretized microstructure can be written as \cite{Cecen2016}

\begin{align} \label{eqn:tps}
    f_t^{\beta \beta^\prime} = \frac{1}{S}\sum_{s=1}^S m_s^\beta m_{s+t}^{\beta^\prime},
\end{align}

\noindent where $m_s^\beta$ is the spatially binned analog of the continuous microstructure function $m(\beta,\mathbf{x})$ introduced in \Cref{sec:intro} and thus describes the probability of finding phase $\beta$ at a spatial bin $s$ \cite{Niezgoda2011}. Correspondingly, 2-point statistics, $f_t^{\beta\beta^\prime}$ quantify the probability of finding phases $\beta$ and $\beta^\prime$ separated by a discretized vector $t$ \cite{Cecen2016}. 

The current MKS approach summarized in \Cref{eqn:poly1,eqn:pca,eqn:tps} allows us to build microstructure-sensitive reduced-order models (or linkages) for the effective composite strength. Furthermore, as shown recently \cite{Latypov2017}, the same approach can be also successfully employed for reduced-order modeling of the  strain-rate partitioning ratios (i.e., $\chi_\beta$). At the same time, as outlined in \Cref{sec:intro}, the current MKS approach needs recalibration of \Cref{eqn:poly1} for every new combination of strengths and strain-hardening laws of the constituents. Indeed, the linkage expressed by \Cref{eqn:poly1} includes only the microstructure terms, $\alpha$, and thus can be calibrated and used only for a specific combination of the constituents with fixed strength and hardening characteristics. This represents the critical gap hindering the development of microstructure-sensitive predictions of the composite stress--strain responses. Here, we pursue a generalization of the MKS approach to account for various combinations of strengths and strain hardening laws for the constituents in the composite. Our main strategy will be to include the strength contrast defined as

\begin{align} \label{eqn:eta}
    \eta = \frac{g_2}{g_1},
\end{align}
into \Cref{eqn:poly1} as a continuous input variable in addition to the microstructure terms, $\alpha$. 

\subsection{Generalized MKS approach incorporating strength contrast} 
\label{sec:mks-new}

Since the strength contrast, $\eta$, is a physics-capturing variable (as opposed to variables $\alpha$ that describe the microstructural topology), the best way to introduce $\eta$ into the linkage in \Cref{eqn:poly1} is to treat the influence coefficients, $A$, as functions of $\eta$. This approach is quite natural as the influence terms, $A$, are closely related to the Green's functions employed in composite theories. Therefore, we replace $A_q$ with $A_q \left (\eta \right)$ in \Cref{eqn:poly1}. Building on prior experience \cite{Gupta2015,Paulson2017,Latypov2017}, we will explore these functional dependencies as polynomial functions. An advantage of using polynomial functions is that it allows merging the polynomial expressions in both $A_q$ and $\alpha^q$ into a single consistent analytical form expressed as

\begin{subequations}\label{eqn:poly2}
\begin{align} \label{eqn:poly2_lnk}
    \bar{p} = \sum \tilde{A}_q \tilde{\alpha}^q (\alpha,\eta), \quad \textnormal{with} 
\end{align}
\begin{equation} \label{eqn:poly2_vars}
    \tilde{\alpha}^q (\alpha,\eta) = \alpha_1^{q_1}\alpha_2^{q_2} \ldots \alpha_R^{q_R}\eta^{q_{R+1}},
\end{equation}
\end{subequations}

\noindent where $\tilde{\alpha}^q$ are now expanded to be power products of PC scores, $\alpha$, and strength contrast, $\eta$. This analytical form of the linkage can be further refined using  physical insights related to $\eta$. Consistently identifying the hard constituent as \texttt{phase 2} and the soft constituent as \texttt{phase 1}, i.e., requiring $g_2 \ge g_1$, we can restrict the range of interest for the contrast ratio as $\eta \in [1,+\infty)$. Furthermore, $\eta=1$ is a special case that corresponds to $g_2=g_1=\bar{g}$ for effective strength and $\dot{\epsilon}_1=\dot{\epsilon}_2=\dot{\bar{\epsilon}}$ for strain rates, which means that the micromechanical response becomes independent of the microstructure and can be thought of as the response of a homogeneous material. One can incorporate this requirement into \Cref{eqn:poly2} by re-expressing the polynomials in terms of $\eta-1$ instead of $\eta$. Finally, it is advantageous to seek responses as ratios, e.g., as effective strength of the composite normalized by the strength of the soft phase, $\bar{g}/g_1$, and as partitioning ratios, i.e., phase-average strain rates normalized by the imposed strain rate, $\chi_\beta$. These ideas lead to the following final forms of the desired linkages:

\begin{subequations}
\label{eqn:mks}

\begin{align} \label{eqn:mks_s}
    \frac{\bar{g}}{g_1} = 1 + \sum \tilde{A}_q^g \tilde{\alpha}^q (\alpha,\eta)
\end{align}

\noindent for the effective strength of the composite (\emph{$\bar{g}$-linkage}) and 

\begin{align} \label{eqn:mks_chi}
    \chi_\beta = 1 + \sum \tilde{A}_q^{\chi_\beta} \tilde{\alpha}^q (\alpha,\eta)
\end{align}

\noindent for strain-rate partitioning ratios (\emph{$\chi$-linkages}), where 

\begin{equation} \label{eqn:mks_vars}
    \tilde{\alpha}^q (\alpha_j,\eta) = \alpha_1^{q_1}\alpha_2^{q_2} \ldots \alpha_R^{q_R}(\eta-1)^{q_{R+1}} \textnormal{ with } q_{R+1} \ne 0.
\end{equation}
\end{subequations}

The expressions in \Cref{eqn:mks} successfully incorporate all the desired physics-based requirements into the linkages. First, they eliminate the terms containing only the PC scores, $\alpha$, without contrast, $\eta$; as explained before, such terms cannot be allowed in the linkages as they must produce the expected trivial solution of uniform fields for the case of $\eta=1$. Indeed, for the special case of $\eta=1$, the linkages produce the expected results of $\chi_1 = \chi_2 = 1$ (i.e., $\dot{\epsilon}_1=\dot{\epsilon}_2=\dot{\bar{\epsilon}}$) and $\bar{g} = g_1 = g_2$, meaning that the effective strength and strain-rate partitioning are independent of the details of the microstructure captured by the PC scores, $\alpha$, when $\eta=1$. Formulation of the new MKS approach in the form shown in \Cref{eqn:mks} is critical for the overall success of the plasticity model presented here, and is one of the important contributions of this work.

The influence coefficients, $\tilde{A}_q^g$ and $\tilde{A}_q^{\chi_\beta}$, in \Cref{eqn:mks} can be obtained by calibration, which requires a one-time effort that  typically involves (i) constitution of an ensemble of digital microstructures (i.e., training set) with a variety of shapes and volume fractions of the constituents, (ii) their quantification by 2-point statistics and PCA, (iii) evaluation of their micromechanical responses by finite element simulations, and (iv) estimation of the influence coefficients by regression methods (on training data) and their verification by new testing data. 

\section{Plasticity Framework Based on the Generalized MKS Approach}
\label{sec:plasticity}

\begin{figure}[!ht]
  \centering
  \includegraphics[width=\textwidth]{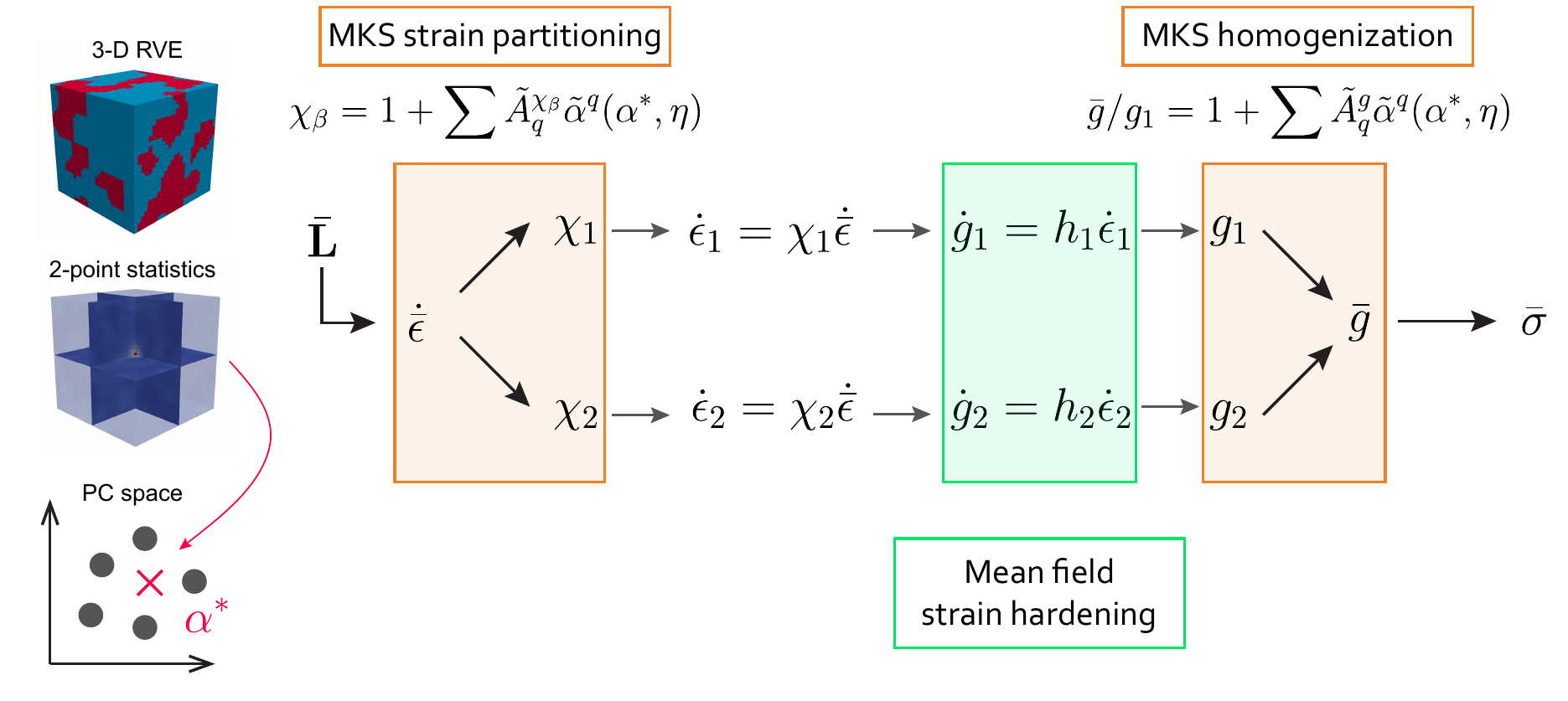}
  \caption{Illustration of the MKS-based plasticity framework.}
  \label{fig:frame}
\end{figure}

The extended MKS approach for capturing the strain-rate partitioning and strength homogenization (\Cref{sec:mks-new}) together with the mean-field approximation (\Cref{sec:mean-field}) allow us to develop a computationally efficient plasticity framework for microstructure-sensitive predictions of stress--strain curves of composites. The plasticity framework proposed in the present work is illustrated in \Cref{fig:frame} and includes the following steps. 

\noindent \emph{Pre-Processing Step}. For the given microstructure of the composite (whose stress--strain curves we wish to predict) compute 2-point statistics and transform them into the PC basis established in the linkage calibration effort. This transformation yields the PC scores, $\alpha^*$ (which we indicate with an asterisk to signify that they represent an RVE different from those included in the training ensemble). Next, for each time increment, $\Delta t= \tau-t$, perform Steps 1--5 below.

\noindent \emph{Step 1}. Obtain the contrast ratio from the current phase strengths (which, at time $\tau=0$, correspond to the initial conditions):

\begin{align} \label{eqn:mdl_eta_t}
    \eta^t = \frac{g^t_2}{g^t_1}.
\end{align}

\noindent \emph{Step 2}. Neglecting the elastic strains, compute the plastic strain rate tensor, $\bar{\mathbf{D}}_p^\tau$, from the imposed velocity gradient tensor, $\bar{\mathbf{L}}$, and its equivalent value as

\begin{subequations}\label{eqn:mdl_strn}
\begin{align} \label{eqn:mdl_strn_tnsr}
    \bar{\mathbf{D}}^\tau_p = \frac{1}{2} \left ( \bar{\mathbf{L}}^\tau + (\bar{\mathbf{L}}^\tau)^T \right),
\end{align}
\begin{align} \label{eqn:mdl_strn_eq}
    \dot{\bar{\epsilon}}^\tau = \sqrt{\frac{2}{3} \norm{\bar{\mathbf{D}}^\tau_p}}.
\end{align}
\end{subequations}

\noindent \emph{Step 3}. Using the calibrated $\chi$-linkages (\Cref{eqn:mks_chi}) for strain-rate partitioning ratios, $\chi_\beta$, partition the imposed equivalent strain rate into each phase taking into account the microstructure (represented by $\alpha^*$) and the current contrast ratio, $\eta^\tau$, and compute the phase-averaged equivalent plastic strain rates, $\dot{\epsilon}_\beta^\tau$, in both phases as

\begin{subequations}\label{eqn:mdl_chi}
\begin{align} \label{eqn:mdl_chi_lnk}
  \chi_\beta^\tau = 1 + \sum \tilde{A}_q^{\chi_\beta} \tilde{\alpha}^q (\alpha^*,\eta^t),
\end{align}
\begin{align} \label{eqn:mdl_chi_def}
  \dot{\epsilon}_\beta^\tau = \dot{\bar{\epsilon}}^\tau\chi_\beta^\tau. 
\end{align}
\end{subequations}

\noindent \emph{Step 4}. Compute the rates of change of strength in the constituent phases, $\dot{g}_\beta$, using the partitioned strain rates,
$\dot{\epsilon}_\beta^\tau$, and the selected hardening laws, $h_\beta$, for both phases using \Cref{eqn:hard_ph}:

\begin{align} \label{eqn:mdl_gevo}
    \dot{g}_\beta^\tau = h_\beta \dot{\epsilon}_\beta^\tau
\end{align}

\noindent and update the strengths of the constituent phases, $g_\beta$, as

\begin{align} \label{eqn:mdl_gupd}
  g_\beta^\tau = g_\beta^t + \dot{g}_\beta^\tau\Delta t.
\end{align}

\noindent \emph{Step 5}. Update the strength contrast ratio based on the new strength values as

\begin{align} \label{eqn:mdl_eta_tau}
    \eta^\tau = \frac{g^\tau_2}{g^\tau_1}
\end{align}

\noindent and compute the effective strength for the composite using the calibrated $\bar{g}$-linkage (\Cref{eqn:mks_s}) for strength, $\bar{g}$, as

\begin{align} \label{eqn:mdl_g_lnk}
  \bar{g}^\tau = g_1^\tau\left(1 + \sum \tilde{A}_q^g \tilde{\alpha}^q (\alpha^*,\eta^\tau) \right ),
\end{align}

\noindent \Cref{eqn:mdl_g_lnk} provides a new value of the effective stress, $\bar{\sigma}$, at the end of the time increment. For rate independent materials, $\bar{\sigma}^\tau = \bar{g}^\tau$.

The underlying assumption of this framework is that 2-point statistics of the microstructure do not change significantly during the imposed overall deformation. This clearly needs to be addressed for large plastic deformations, and is a focus of ongoing development in this field.

We note that the present approach employs a similar strategy as the prior work by Liu et al.\ \cite{Liu2016} in learning the underlying physics from numerical simulations. At the same time, there are significant conceptual differences between the two approaches. In the Liu et al. [53] approach, clusters of material points that are alike in terms of localizing elastic strain are identified by $k$-means clustering of concentration tensor field and are treated as distinct phases in a self-consistent homogenization scheme for inelastic behavior. It is implicitly assumed that localization in the elastic regime also holds during inelastic deformation. In contrast, the present approach learns salient information on strain-rate partitioning and homogenization directly from the numerical simulations that accurately account for the inelastic phases. Furthermore, unlike the self-consistent homogenization scheme, our approach does not require the user to select the properties of a reference medium, which simplifies the prediction of the stress--strain responses for new microstructures.

\section{Case Study}
\label{sec:case}

In this section, we present in detail an application of the new microstructure-sensitive reduced-order modeling framework developed in \Cref{sec:mdl-dev,sec:plasticity}. The specific goal is the prediction of tensile stress--strain curves of composites with two isotropic phases exhibiting linear and nonlinear strain hardening laws. As pointed out above, the workflow of the linkage calibration effort includes (i) constitution of a training ensemble of digital microstructures, (ii) their quantification by 2-point statistics and PCA, (iii) estimation of their micromechanical responses by finite element simulations, and (iv) calibration and verification of the MKS linkages. Upon calibration, the homogenization and strain-rate partitioning linkages can be used in the plasticity framework presented in \Cref{sec:plasticity} for predictions of the stress--strain curves. 

\subsection{Calibration ensemble of digital microstructures}
\label{sec:ms-gen}

\begin{figure}[!ht]
  \centering
  \includegraphics[width=\textwidth]{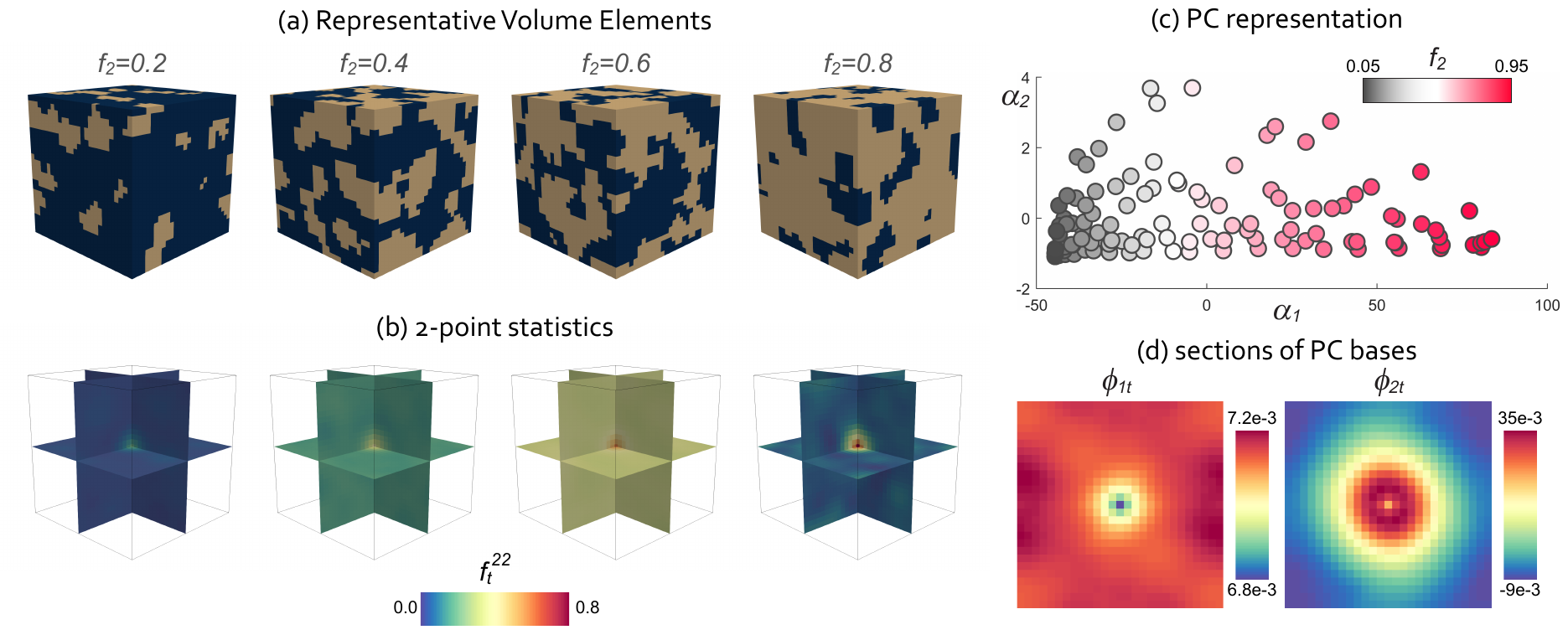}
  \caption{Microstructures generated in Dream.3D and their quantification: (a) exemplar RVEs of the ensemble and (b) their 2-point statistics shown on orthogonal middle sections; (c) 2-D projection of low-dimensional representation of 114 microstructures obtained by PCA of 2-point statistics; (d) 2-D middle sections of the first two PC basis (normal to $x$ axis).}
  \label{fig:ms}
\end{figure}

The primary goal of this step is to generate a rich selection of digital microstructures with a wide range of volume fractions and diverse spatial configurations of the constituents to build broadly applicable linkages for strain-rate partitioning and strength homogenization (\Cref{eqn:mks}). In this case study, a calibration set of microstructures with various phase topologies was generated with open-source software Dream.3D \cite{Groeber2014}. The microstructures were generated as periodic RVEs using six different sets of particle size characteristics (parameters employed in Dream.3D \cite{Groeber2014} for the microstructure generation), and volume fractions of the constituent phases were varied systematically from 0.05 to 0.95, with a 0.05 step size. The RVE size was selected to ensure that it was sufficiently large to capture the length scale of interaction between the constituents, while also keeping it small enough so as to not incur a very large computational cost. The first requirement is fulfilled if the Green's function underlying the microstructure--property relationship approaches values sufficiently close to zero within half of the RVE size. Therefore, localization kernels related to the Green's function \cite{Yabansu2014} and their characteristic decay length scale should be used to guide the selection of the RVE size. Our previous work on  localization and homogenization problems \cite{Yabansu2014,Paulson2017} suggests that this requirement can be satisfied in $21^3$ RVEs for composites with low to moderate contrasts. In this study, we set the RVE size to $27 \times 27 \times 27$ voxels, which is slightly larger than the RVE size in prior work because of the wider range of strength contrasts considered. Besides the RVE size, special consideration is also required in the selection of the optimal number of RVEs in the calibration ensemble. The goal here is to make a trade-off such that the number of RVEs is large enough to provide data needed to calibrate robust and broadly applicable linkages, while keeping reasonable the one-time computational cost of FE simulations for the entire ensemble of RVEs. Since rigorous protocols for selecting the optimal number and types of RVEs for building reduced-order models in the MKS framework are currently lacking, it was determined by trial-and-error for the present study. Specifically, the minimal number of RVEs -- 114 -- which led to small errors in cross-validation (see below) was adopted for the present case study. Since each microstructure is used with nine different values of contrast ratios, this amounts to a total of 1026 FE simulations. Examples of the 114 two-phase 3-D RVEs produced for this study are shown in \Cref{fig:ms}a.

\subsection{Quantification of microstructures}
\label{sec:ms-quant}

The generated microstructures were first quantified by 2-point statistics with the aid of computationally efficient DFT-based approaches \cite{Cecen2016}. 
 Since the microstructures used in the study are three-dimensional, their corresponding directionally resolved 2-point statistics are also three-dimensional (see examples in \Cref{fig:ms}b). 

PCA was performed on the set of 2-point statistics for obtaining low-dimensional representations of the microstructures in the ensemble, while also establishing the basis for the consideration of other new microstructures for which predictions of the stress--strain curves will be sought. A projection of the low-dimensional representation of 114 microstructures is shown in \Cref{fig:ms}c in terms of the first two PCs ($\alpha_1$ and $\alpha_2$); the  corresponding basis maps are visualized in \Cref{fig:ms}d as 2-D middle sections. The ordered orthogonal decomposition into PC scores and PC bases provided by PCA allows for the use of just a handful of scalar PC scores in establishing efficient linkages \cite{Gupta2015,Latypov2017,Paulson2017}. Furthermore, retaining the PC bases enables interpretation of the PC scores, transformations of new microstructures into the PC space, as well as reconstructions of the original 2-point statistics to a good approximation. For example, in PCA applied on the present ensemble of 2-point statistics, the first PC captures more than \SI[mode=text]{99}{\percent} variance, which is typical for ensembles with a wide range of volume fractions of the constituents \cite{Latypov2017,Cecen2014}. Basis maps corresponding to the first two PCs (represented by 2-D sections) exemplify the complex patterns in respect to which the PC scores capture differences in 2-point statistics in the ensemble of the microstructures. While only two first PCs are shown in \Cref{fig:ms}, a wider range of PCs was considered for calibration of the linkages as described in \Cref{sec:cal}.

\subsection{Numerical simulations of micromechanical responses}
\label{sec:fea}

The micromechanical responses of all the generated microstructures were simulated using the commercial finite element (FE) software ABAQUS \cite{hibbett1998abaqus}. Each cuboidal voxel of the digital two-phase microstructure was converted into a C3D8 element \cite{hibbett1998abaqus}. Equal-volume cuboidal elements were chosen for reducing the computational cost involved in learning the desired linkages. This is because these meshes allow direct application of DFT representations of the microstructure and FFT algorithms, as explained in \Cref{sec:ms-quant}. In the FE models, the behavior of each phase was assumed to be governed by an elastic--plastic isotropic J$_2$ plasticity model based on von Mises yield criterion and the rate-independent associated flow rule. To focus on plastic behavior of the composite RVEs, the elastic strains were minimized by adopting large values of elastic moduli for both phases: $g_2/E_2=g_1/E_1 = 10^{-6}$. A range of strength contrasts, $\eta \in [2,10]$, was considered and implemented in FE models by setting the strength of the soft phase to 
1, i.e., $g_1=1$ (arbitrary units, a.u.), and that of the hard phase to the value of the contrast of interest, i.e., $g_2=\eta g_1$, where $\eta \in \{2,3,\ldots,10 \}$. The resulting $9 \times 114$ 3-D FE meshes were subjected to an isochoric tensile deformation described by the following components of the imposed velocity gradient tensor $\bar{L}_{33} = \bar{L}_{22} = -\bar{L}_{11}/2$ with $\bar{L}_{11} = 10^{-3}$ $s^{-1}$ and off-diagonal components being zero. The simulations were carried out for deformation of the microstructures up to an equivalent strain of $\bar{\epsilon} = 0.002$. The targeted deformation was achieved by prescribing the velocity boundary conditions on the corner nodes of the microstructures, while other surface nodes were constrained to produce periodic boundary conditions. Periodic boundary conditions allow the treatment of the microstructure as an infinite field without any explicit boundaries. The use of any other type of boundary conditions generally affects the stress and strain fields near the boundaries of the RVEs differently than the fields inside the RVEs. Periodicity was implemented by pairing displacements of the corresponding nodes on the opposite surfaces using ABAQUS Equation constraint. Simulation results were post-processed to extract the effective strength, $\bar{g}$, and the strain-rate partitioning ratios, $\chi_\beta$. The effective strength of the composite was obtained by first averaging the stress tensor components over all integration points of the RVE, which corresponds to volume-averaging as all elements had the same volume. The volume-averaged stress tensor is equal to macroscopic stress tensor according to Hill--Mandel principle \cite{Hill1963,Mandel1966}. The macroscopic equivalent stress was then computed using the von Mises definition on the averaged stress components, which gives the effective strength of the composite as $\bar{g} = \bar{\sigma}$ for rate-independent plasticity. Strain-rate partitioning ratios, $\chi_\beta$ were computed by averaging the equivalent plastic strain rates over the integration points in the $\beta^\text{th}$ phase and then dividing by the macroscale effective plastic strain rate (computed based on the imposed velocity gradient tensor) according to \Cref{eqn:eps_ph,eqn:chi}. 

The data obtained from these \(9\times 114\) FE simulations was complemented with \emph{a priori} known responses for the trivial case of \(\eta=1\) (\(\chi_1=\chi_2=1\), \(\bar{g}=g_2=g_1\) for all RVEs), which altogether yielded 1140 triplets of $\bar{g}$, $\chi_1$, $\chi_2$.

\subsection{Calibration of MKS linkages}
\label{sec:cal}

The MKS linkages for strain-partitioning and effective strength (\Cref{eqn:mks}) were calibrated using multivariate regression. Multivariate regression allows establishing the unknown coefficients, $\tilde{A}$, of the polynomial terms (\Cref{eqn:mks}) by minimizing the sum of the squares of errors between the linkage and the FE data points \cite{James2013}. 

Calibration of linkages using multivariate regression involves selection of linkage hyper-parameters -- the number of the PC scores to include in the polynomial function of the linkage (i.e., truncation level $R$ in \Cref{eqn:pca}) and the degree of the polynomial, Q. The first $R$ PC scores are typically selected as the ones capturing the most variance in 2-point statistics of the ensemble, which is anticipated to correspond to the variance in the micromechanical responses \cite{Niezgoda2013}. As such, the truncation level, $R$, in \Cref{eqn:pca} can be selected based on the desired level of the cumulative variance in 2-point statistics explained by these $R$ PCs (e.g., close to \SI[mode=text]{100}{\percent}). In addition to the number of PC scores, the degree of the polynomial, Q, also requires careful selection to ensure a good fit and at the same time to avoid overfit. Preventing the latter refers to ensuring the ability of the model to provide accurate predictions on new microstructures outside of the training dataset. The hyper-parameters were determined by systematic trial-and-error search using goodness of fit and cross-validation error measures as the criteria. Furthermore, when the potential polynomial functions were identified, we performed correlation analysis -- Pearson test \cite{Pearson1895} -- between each individual polynomial term (i.e., monomial) and the micromechanical response of interest. These tests were carried out to exclude insignificant terms from the functions, thereby simplifying the linkages and further reducing the risk of overfitting.

This procedure resulted in linkages containing 15, 78, and 71 polynomial products of the PC scores and the strength contrast (with a maximum power of $Q=5$) for the effective strength, $\bar{g}$, and the partitioning ratios, $\chi_1$ and $\chi_2$, respectively. These specific analytical forms of the linkages were found to give the best combination of the goodness of the fit and the cross-validation errors. More complex analytical forms for the $\chi$-linkages (i.e., containing greater number of terms) is consistent with our prior observations \cite{Gupta2015,Latypov2017} that strain-rate partitioning or localization is more challenging  for linkage development compared to effective properties such as strength of the composite. Once the analytical forms were established, the linkages were calibrated by final multivariate regression.  

\begin{figure}[!ht]
  \centering
  \includegraphics[width=\textwidth]{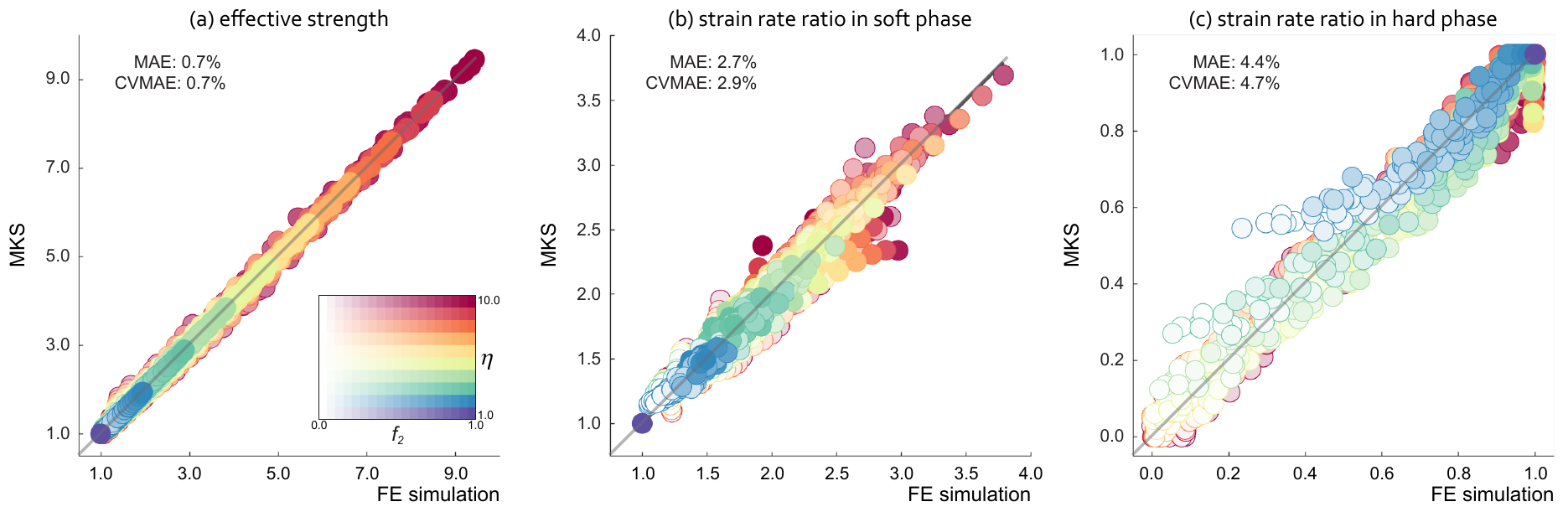}
  \caption{Calibration results shown as parity plots for (a) effective strength; (b--c) strain-rate partitioning ratios. The data points are color coded according to the hardness contrast (from violet to red) with color intensity corresponding to the volume fraction of the hard phase (as shown in the inset). Mean absolute errors (MAE) and cross-validation mean absolute errors (CVMAE) are also shown for each case.}
  \label{fig:cal}
\end{figure}

\Cref{fig:cal} shows the results of calibration of all three linkages, for the effective strength, $\bar{g}$, and the partitioning ratios, $\chi_\beta$, performed on the data obtained from FE simulations for the entire range of strength contrasts, $\eta \in [1,10]$. The results are visualized as parity plots between the FE simulation results and the linkage calibration. In these plots, points lying along the parity line indicate a perfect fit. The parity plots illustrate that reasonable fits were obtained for all three micromechanical responses. The reasonable fits are also seen in the low values of the mean absolute errors (MAE): \SI[mode=text]{0.7}{\percent} for the $\bar{g}$-linkage, $\sim$ 3--\SI[mode=text]{5}{\percent} for the $\chi$-linkages. 

In addition to the goodness of fit, which shows how well the calibration data can be fitted to the chosen function, it is also important to perform cross validation of the calibrated models, which characterizes their ability to generalize and thus their actual predicative power. Indeed, functions overfitted to the calibration dataset will perform badly on new microstructures. In prior work \cite{Gupta2015,Latypov2017,Paulson2017}, leave-one-out cross validation (LOOCV) proved to be an effective technique to evaluate overfits, especially in the context of micromechanics when the calibration data is often limited because of the computational expenses associated with generating the calibration data (with FE simulations). LOOCV tests the sensitivity of the calibrated functions to exclusion of a single data point from calibration: high sensitivity indicates overfit of the functions, while well fitted functions are insensitive to leaving out only one sample. LOOCV performed on the three linkages showed that LOOCV errors are only slightly higher than the corresponding MAE, which indicates that the linkages can be expected to generalize well to microstructures outside of the calibration ensemble.

\subsection{Verification of MKS linkages}
\label{sec:val}

To further verify and critically evaluate the linkages calibrated in \Cref{sec:cal}, we compared their predictions against FE results on new microstructures and strength contrasts distinct from those included in the calibration dataset. Such verification can be viewed as testing the interpolation capabilities of the calibrated linkages in the microstructure--contrast space. In this work, we carried out three separate verification exercises involving (i) new microstructures with contrasts used in the calibration, (ii) new contrasts with microstructures used in the calibration, and (iii) new microstructures and new contrasts. For verification tests (i) and (iii) that require new microstructures, a new set of RVEs was generated using Dream.3D with inputs different from those utilized for the generation of the training RVEs. Specifically, 38 RVEs with two sets of new size and shape characteristics were obtained covering the volume fraction range from 0.05 to 0.95 of the hard phase. For verification test (ii), a subset of 19 RVEs was sampled from the calibration ensemble. All verification RVEs were quantified by 2-point statistics (as described in \Cref{sec:ms-quant}) followed by their transformation into the PC space already established in the calibration step. In other words, the new RVEs generated for the verification were projected into the PC space identified using only the calibration RVEs. The PC scores were used together with contrast values for the predictions of effective strength, $\bar{g}$, and strain-rate partitioning ratios, $\chi_\beta$ using the corresponding linkages obtained in \Cref{sec:cal}. To verify these predictions, the corresponding FE simulations were performed following the protocols described in \Cref{sec:fea}.

\begin{figure}[!ht]
  \centering
  \includegraphics[width=\textwidth]{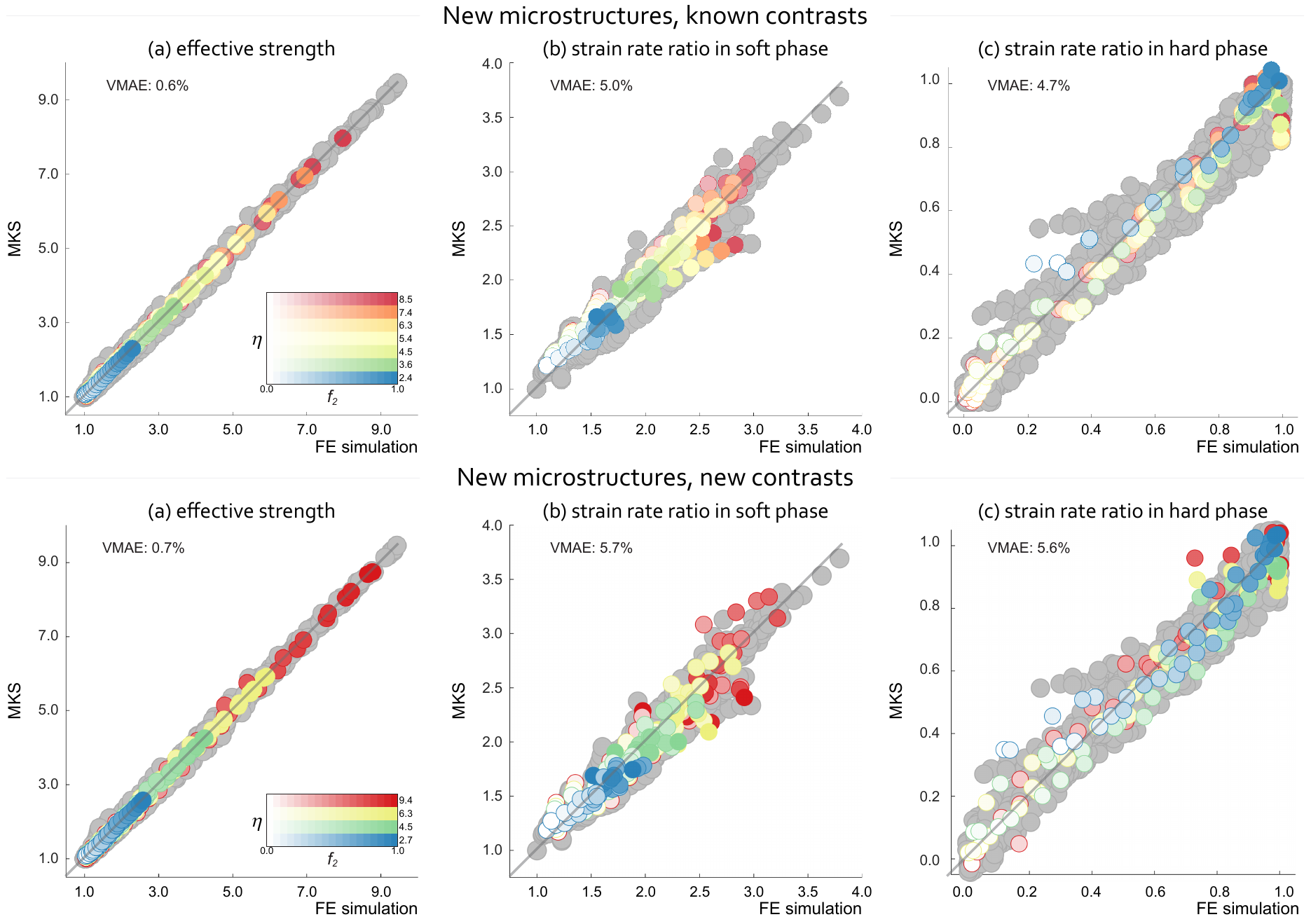}
  \caption{Verification results for interpolation tests shown as parity plots for (a) effective strength; (b--c) strain-rate partitioning ratios. The data points are color coded according to the hardness contrast (from blue to red) with color intensity corresponding to the volume fraction of the hard phase (as shown in the inset). Calibration data are shown in gray in the background. Validation mean absolute errors (VMAE) are also shown for each case.}
  \label{fig:val}
\end{figure}

\Cref{fig:val} presents the results of the verification tests as parity plots for predictions of effective strength and strain-rate partitioning ratios\footnote{results of test (i) are not shown being very similar to the presented parity plots and within the indicated error ranges}. It is seen that the predictions for the verification tests lie within $\sim$ \SI[mode=text]{1}{\percent} MAE for the effective strength and within $\sim$ \SI[mode=text]{6}{\percent} MAE range for the partitioning ratios (errors are defined with respect to the FE simulation results). These verification results demonstrate that the MKS linkages can be used for accurate predictions of strain-rate partitioning ratios and effective strength for a variety of combinations of two-phase microstructures and strength contrasts. These features make the homogenization and strain-rate partitioning linkages suitable for incorporation into the plasticity model outlined in \Cref{sec:plasticity} for microstructure-sensitive predictions of stress--strain curves, which is tested in the next Section. 

\subsection{Prediction of stress--strain curves}
\label{sec:ss-curves}

In this section, we implement the calibrated and verified MKS linkages into the plasticity model presented in \Cref{sec:plasticity}. One of the advantages of the present approach is that the development of MKS linkages presented above is completely independent of the selection of the hardening laws capturing the strain hardening behavior of the constituents in the composite, for which we seek to predict the effective stress--strain curve. It means the MKS linkages that we calibrated and verified in the previous sections can be used for virtually any combination of the hardening laws of the constituents (without recalibration) as long as the strength contrast during deformation of the composite is within the range of contrasts used in the calibration, i.e., $\eta \in [1,10]$. 

\begin{figure}[!ht]
  \centering
  \includegraphics[width=0.76\textwidth]{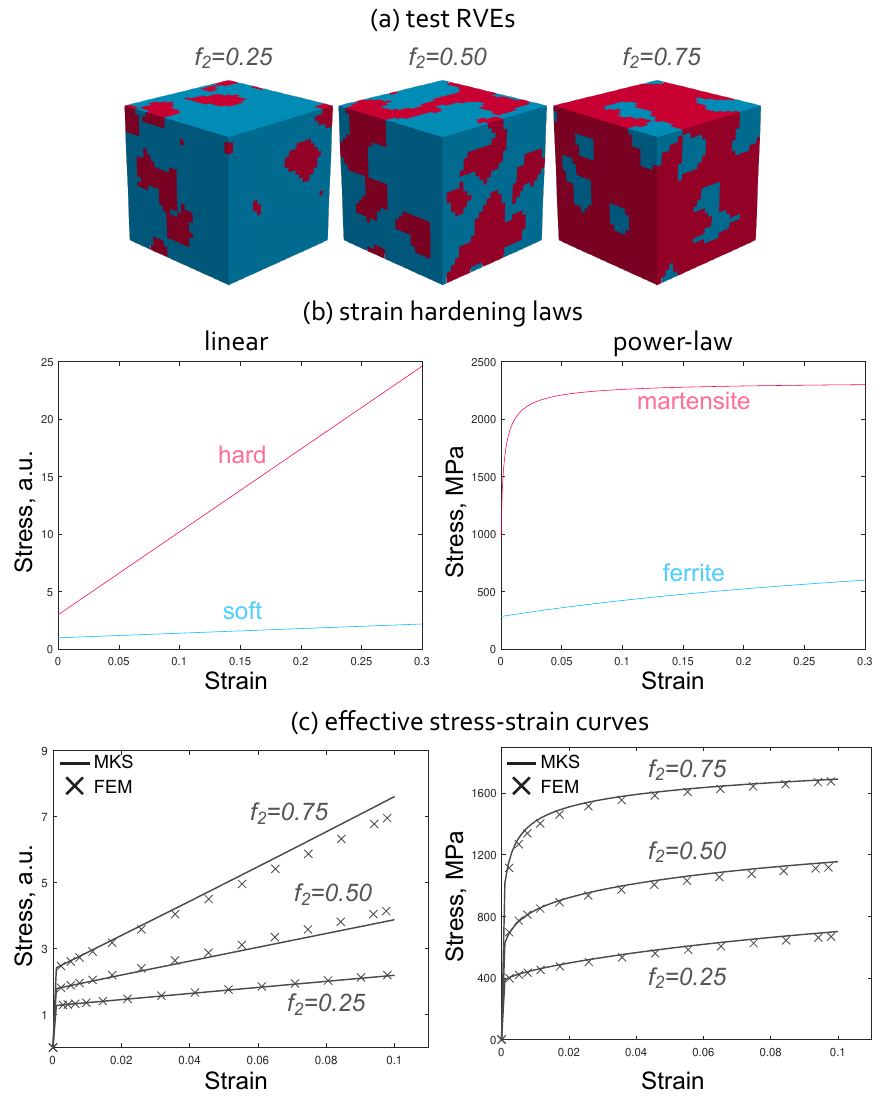}
  \caption{Evaluation of the MKS-based model for the prediction of effective stress--strain curves: (a) test RVEs, (b) linear and power-law strain hardening laws of the constituents, (c) predicted stress--strain curves for the three RVEs by MKS-based model and FEM for the case of linear strain hardening laws and power-law strain hardening of ferrite and martensite.}
  \label{fig:ss}
\end{figure}

To test this capability of the MKS-based plasticity model, we evaluate the predictions of the stress--strain curves of new RVEs (``unseen'' during calibration) with linear and nonlinear hardening laws of the constituents. These predictions are tested against more computationally demanding FE simulations of tensile deformation of the same RVEs. FE simulations were carried out using the protocols described in \Cref{sec:fea} with the only difference that the simulations for obtaining stress--strain curves were performed to larger strain, $\bar{\epsilon} = 0.1$. The three test RVEs, for which stress--strain curves were obtained with the MKS-based plasticity model and the FE simulations, are shown in \Cref{fig:ss}a. 

\paragraph{Linear hardening laws}  In the first test case, we compare the new plasticity model against FE simulation results for composites with both phases exhibiting linear hardening laws. For linear hardening laws, \Cref{eqn:hard_ph} becomes

\begin{align} \label{eqn:hard_linear}
    \dot{g}_\beta = h^c_\beta \dot{\epsilon}_\beta, \quad  h^c_\beta = \text{const}
\end{align}

The specific values adopted for modeling were $h^c_1 = 4$ and $h^c_2 = 72$ with initial slip resistances assigned as $g_1 = 1$ and $g_2 = 3$. These strain hardening laws are visualized in \Cref{fig:ss}b. 
 
\paragraph{Nonlinear hardening laws} In the second case, the plasticity model was tested on composites with both phases following nonlinear hardening laws. The selection of models and hardening parameters was based on data reported previously for a dual-phase steel \cite{Tasan2014a}. Specifically, we consider power-law strain hardening model, for which \Cref{eqn:hard_ph} is expressed as follows \cite{Peirce1983}: 

\begin{align} \label{eqn:hard_power}
    \dot{g}_\beta = h^0_\beta \left ( 1 - \frac{g_\beta}{g_\beta^\infty}\right) ^ {a_\beta} \dot{\epsilon}_\beta,
\end{align}

\noindent where $\{h^0_\beta, g_\beta^\infty, a_\beta \}$ are phase-specific sets of hardening parameters. The values for these hardening parameters were selected to capture strain hardening laws reported for ferrite and martensite in dual phase steel \cite{Tasan2014a}: $g_1 = $\SI[mode=text]{285}{MPa}, $g^\infty_1 = $\SI[mode=text]{1236}{MPa}, $h^0_1 = $ \SI[mode=text]{3.0}{GPa} for ferrite and $g_2 =$ \SI[mode=text]{974.4}{MPa}, $g^\infty_2 =$ \SI[mode=text]{2330.4}{MPa}, $h^0_2 =$ \SI[mode=text]{1351.2}{GPa}  for martensite with $a = 2.25$ for both phases (the corresponding curves are also shown in \Cref{fig:ss}b.). 

\Cref{fig:ss}c shows the stress--strain curves predicted by the present model for three RVEs with different phase topologies and volume fractions (\SI[mode=text]{25}{\percent}, \SI[mode=text]{50}{\percent}, and \SI[mode=text]{75}{\percent} of the reinforcement phase) as well as two cases of strain hardening laws of the constituents (linear and power-law). The results from FE simulations obtained for the same RVEs are also shown in \Cref{fig:ss}c. It is seen that the present model compares quite well with the FE results. 

\section{Summary}
\label{summary}

In the present study, we have developed a new Materials Knowledge System framework for microstructure-sensitive predictions of stress--strain relationships in composites with isotropic constituents. The model is developed to cover a wide range of combinations of the phase topologies, volume fractions, and strain hardening laws. Such versatility of the model relies on the three key components: (i) mean-field approximation of strain localization, strength homogenization, and strain hardening, (ii) MKS for strain-rate partitioning, (iii) MKS for strength homogenization. To enable incorporation of diverse strain hardening laws of individual constituents of composite materials, the MKS approach to strain-rate partitioning and homogenization was generalized in the present work to include strength contrast as a continuous variable alongside microstructure terms in the MKS microstructure--response linkages. Although, there is no theoretical limitation on the particle shapes that can be included in the calibration of the linkages, some practical limitations do arise from the finite element meshing scheme adopted in generating the numerical simulation datasets from which the linkages are learned.  In this work, a uniform mesh of cuboidal elements was employed for generating all of the numerical simulation datasets. This was done mainly to simplify the computations involved in the learning of the linkages (using the FFT algorithms). In future work, extensions to this framework may be sought to get around this limitation.

Upon the calibration of the linkages, the present MKS model showed good agreement with full-field FE simulations in prediction of effective stress--strain curves, offering, at the same time, significant savings of the computational resources. Indeed, the CPU time for FE simulations was $\sim$ \SI[mode=text]{10}{min} for a single RVE, whereas the same prediction of the stress--strain curves using the present model took only $\sim$ \SI[mode=text]{0.5}{s}. Furthermore, the present model does not require any FE software and can be implemented in any of the widely used scripting languages, such as MATLAB or Python. Example MATLAB scripts demonstrating the implementation of the calibrated linkages (along with the necessary data) and their use for predicting stress--strain responses are made available to the community via GitHub.\footnote{Repository address: \href{https://github.com/mined-gatech/MKS-hardening}{https://github.com/mined-gatech/MKS-hardening}} The versatility and the computational efficiency of the model therefore makes it suitable for incorporation into emerging cloud-based ICME platforms as a powerful tool for rapid screening of potential microstructures for performance optimization in materials design \cite{Latypov2019}.

We finally note that the present work is the first demonstration of predictive capabilities of the data-driven models based on the MKS approach to micromechanics beyond the elastic and initial yield regimes of deformation pursued in previous studies. This work therefore represents a major advance towards computationally efficient multiscale modeling and design of materials.

\section*{Acknowledgments}

MIL and LST acknowledge the support by the French State through the program ``Investment in the Future'' operated by the French National Research Agency (ANR) and referenced by ANR-11-LABX-0008-01 (LabEx DAMAS). MIL also acknowledges the support from National Science Foundation grant SI2-SSI:LIMPID, NSF \#1664172. SRK acknowledges the support from ONR grant N00014-15-1-2478.

\section*{References}
\bibliography{refs.bib}{}
\bibliographystyle{elsarticle-num}

\end{document}